\documentclass[useAMS,usenatbib]{mn2e}
\usepackage{amsfonts}
\usepackage{amssymb}
\usepackage{bbm}
\usepackage{graphicx}
\usepackage{color}

\title[CS 5-4  in galaxies]
{CS 5-4 survey toward nearby IR bright galaxies}
\author[J. Wang et al.]
{Junzhi Wang$^{1,2,3}$\thanks{email: junzhiwang@nju.edu.cn}, Zhiyu Zhang$^{3,4,5}$,  Yong Shi$^6$  \\
$^1$Department of Astronomy, Nanjing University, 22
Hankou Road, Nanjing, 210093, China\\
$^2$Key Laboratory of Modern Astronomy and  Astrophysics (Nanjing University), Ministry of Education, Nanjing 210093, China\\
$^3$Purple
Mountain Observatory, CAS, 2 West Beijing Road, Nanjing, 210008,
China\\
$^4$Max-Planck-Institut f{\"u}r Radioastronomie, Auf dem H\"ugel 69,
D-53121 Bonn, Germany\\
$^5$Graduate School of the Chinese Academy of Sciences, 19A Yuquan Road,
P.O. Box 3908, Beijing 100039, China\\
$^6$Infrared Processing and Analysis Center, California Institute of Technology, 1200 E. California Blvd, Pasadena, CA 91125}
\begin{document}


\pagerange{\pageref{firstpage}--\pageref{lastpage}} 

\maketitle

\label{firstpage}

\begin{abstract}

With the observations of the  CS 5-4 line toward  a sample of 24 infrared bright galaxies using HHSMT,   we detected 
 CS 5-4 emission in 14 galaxies, including 12 ULIRGs/LIRGs and 2 nearby normal galaxies.  As a good dense gas tracer, which has been well used for studying star formation in the Milky Way,  CS 5-4  can trace the active star forming gas in galaxies. The correlation  between CS 5-4 luminosity, which is estimated with  detected CS 5-4 line emission, and infrared luminosity in these 14
 galaxies is fitted  with correlation coefficient of   0.939 and the slope close to unity.    This correlation confirmed that dense gas, which is closely linked to star formation, is very important for understanding star formation in galaxies.
\end{abstract}


\section{Introduction}
  The star formation law, the relation between star formation rate (SFR)
 and local gas density, is a fundamental tool to study star formation processes and understand galaxy evolution.
This relation was first proposed by  \cite{1959ApJ...129..243S}  as
$\rho\rm_{SFR} \sim (\rho\rm_{gas})$$^N$, where $\rho\rm_{SFR} $ and $\rho\rm_{gas}$ are the volume densities of SFR and gas, respectively, while  $N \approx$  2 and the
atomic gas density was used for discussion. With measurements of disk-averaged total gas surface density $\Sigma\rm_{gas}
= \Sigma\rm_{HI} + \Sigma\rm_{H_2}$ and the 
disk-averaged star formation surface density $ \Sigma\rm_{SFR}$  
   in 61 normal
spiral galaxies and  36 infrared-selected
starburst galaxies, \cite{1998ApJ...498..541K} gave the best fit
for the  Schmidt law with index $N=1.4\pm0.15$.  On the other hand, with pixel-by-pixel analysis at sub-kpc scales of 18 nearby galaxies, which include 7 spiral galaxies and 11  late-type/dwarf galaxies,   \cite{2008AJ....136.2846B} found that a best-fit molecular Schmidt law of  $ \Sigma\rm_{SFR}= 10^{-2.1\pm0.2} \Sigma\rm_{H_2}^{1.0\pm0.2}$. Similar results ($N\sim1$) were obtained with the data of 23 nearby galaxies \citep{2008AJ....136.2782L},  while the slope changes to 1.17  using galaxies including  $z \sim 1-3$ normal  star-forming galaxies  \citep{2010MNRAS.407.2091G}.  But if the   ultraluminous infrared galaxies (ULIRGs)  with extremely high star formation rate, which  have the star formation efficiency  four to 10 times than those normal star forming galaxies \citep{2010MNRAS.407.2091G}, are included, the slope will be higher than the results from \cite{2008AJ....136.2846B}, \cite{2008AJ....136.2782L} and \cite{2010MNRAS.407.2091G}.

Lines of molecules with high dipole momentum, such as HCN, HNC, CN, HCO$^+$ and CS,  are thought to be better tracers of dense star forming gas than CO lines.   These lines have been used    to  study star formation in the Milky Way, while they are generally difficult to be detected in extragalactic objects due to their faint emission.  HCN, HNC and HCO$^+$ lines 
 \citep{2008A&A...479..703G, 2008ApJ...681L..73B}  have been detected in  a few  galaxies with millimeter (IRAM 30m) and sub-millimeter telescopes (JCMT, HHSMT).  CN line in several galaxies \citep{2002A&A...381..783A}
 also has been observed with SEST.   With the observations of   dense gas tracers toward 65 galaxies  \citep{2004ApJS..152...63G},  which is the largest extra-galatic sample with dense gas tracer observation up to now,  \cite{2004ApJ...606..271G}
 found a strong linear correlation between HCN and IR
luminosity ($N\sim$1)  with  more than 3 
orders of infrared luminosity range, including  LIRGs and ULIRGs.  This relation has been extended to the
Milky Way dense cores \citep{2005ApJ...635L.173W}  and possibly  high-$z$
galaxies and  QSOs as well \citep{2007ApJ...660L..93G}.  Such results  agree well with  that  dense molecular 
gas  is more directly related to final star formation than
total molecular gas traced by CO lines   from local universe  to high-$z$. 

  The line ratios  of HCN,
HNC, CN, and HCO$^+$ molecules  vary in different galaxies due to
different chemical and physical conditions,  and also maybe due to pumping
mechanism, which  means  the dense gas mass estimated 
with the observation of only one dense gas tracer may be biased.
As a good dense gas tracer, CS  has been widely used  to study  dense cores in GMCs in the
Milky Way.   Since it is not affected by shock chemistry, which is the case for HCO$^+$ molecule, and it is not related to HCN, HNC, and CN chemistry,  CS can give an independed view of dense gas in galaxies from those based on 
HCN, HNC, CN, and HCO$^+$.   With  critical density of about  10 times
 that of HCN 1-0 \citep{1999ARA&A..37..311E},   CS 5-4  can trace dense gas in active star forming regions.  CS 5-4 line has been detected in three  nearby galaxies (IC342, NGC 253, and M82)\citep{1989A&A...223...79M},  one nearby major merger NGC 4038/4039 (the Antennae galaxy)  \citep{2008ApJ...685L..35B},   and the nearest ULIRG Arp 220  \citep{2009ApJ...692.1432G}  with JCMT.  CS 5-4 in several galaxies, including sources have been observed with other telescopes listed before and  NGC 1068  \& Maffei 2 have been detected with IRAM 30M \citep{2009ApJ...694..610M}.  Most of such galaxies are  not ULIRGs/LIRGs, which are galaxies with the most active star formation in local universe. 
  So with  observations of CS 5-4 toward a sample of galaxies with more total sources and  including more ULIRGs/LIRGs, we can see if dense gas traced by different  molecules can give  similar correlation or not, by comparing with existing HCN results \citep{2004ApJ...606..271G}.

In this letter, we  describe the observations and data reduction
in \S2, present our main results and discussion in \S3,   and give a brief summary in \S4.

\section{Observations and data reduction}

Our sample is composed of 24 galaxies including 20 local (U)LIRGs
and 4 nearby normal galaxies that have strong CO emissions.
  The (U)LIRGs  are   selected from RBGS \citep{2003AJ....126.1607S} with  $\delta >$ -20$\degr$,  $L_{IR} >$ 10$^{11}$ $L_{\odot}$, $F_{100\mu m} >$ 25 Jy, 
$F_{60 \mu m}>$ 20 Jy. 
   The nearby galaxies in our sample  have been observed with 
millimeter telescopes for HCN 1-0   \citep{2004ApJ...606..271G}, which provides  enough dense molecular gas  traced by HCN 1-0, for higher possibility of CS 5-4 detection.  Table 1 listed the information of sources  in this sample, including  source name, the coordinates (J2000), luminosity distance,   red-shift, and also the infrared luminosity.  The sources in Table 1 are ordered with the  luminosity distance. 

\subsection{CS 5-4 observations with  HHSMT }

The 10 m Heinrich Hertz Submillimeter Telescope (HHSMT) on Mt. Graham, Arizona, 
was used to observe the CS 5-4 line in this sample. These observations 
were done in good ($\tau \le$ 0.3) to medium ($\tau \sim$ 0.3-0.6) 
weather conditions, with typical system temperature of 210 K,  during several runs on the HHSMT in  February 2009, December 2009,
and  February 2010. The 1.3 mm ALMA receiver was employed with
lower side-band tuning to the CS 5-4 frequency in the single-sideband (SSB) mode. The beam size at the frequency for CS 5-4 ($\sim 244GHz$) is $\sim33''$.
In order to ensure the baselines as flat as possible, most observations were done in
fast beam-switching mode with a chop throw of 2$'$ in azimuth (AZ) in a chopping 
frequency of 2.2 Hz. For a few nearby galaxies,   position switching 
mode was used to avoid the background contamination. During the 2010 February run,
the ALMA receiver had only H-pol working properly, while the dual polarization mode was used on other time.

 We used both  the Acoustic Optical Spectrometer (AOS) and Forbes 
Filter Banks (FFB)  simultaneously, with  $\sim$ 1 GHz bandwidth  and 1024 channels
to cover the full velocity-width of the expected broad lines. The data of both spectrometers were used to check with each other, but  not combined together. Most of the final spectra were used the FFB data, for its relative 
stability. Our observing strategy involved pointing 
and calibration observations every 2 hr using Saturn, Jupiter or Venus when available. 
Calibration scans were obtained in position-switching mode with reference position of 
5$\arcmin$ in RA. We found typical pointing errors of 2-3$\arcsec$ and
measured the main-beam efficiency, $\eta_{mb}$, to be 0.6 for filter-bank A 
(H-polarization) and 0.8 for filter-bank B(V-polarization).  
The systematic flux calibration uncertainty is about   20\%.

\subsection{CS data reduction}
All CS data reduction was performed using the CLASS program of the 
GILDAS package developed at IRAM. The  velocity intervals based on 
the HCN or CO line profiles in literature are used to set the velocity 
range for all sources. Each baseline was inspected by eye and quantified 
into different levels by their baseline flatness, system temperature, 
standing wave, and etc. 
About 10\% -15\% bad spectra were discarded from the qualification.   
For each source, the spectra    were  co-added, weighted by the rms noise of 
each spectrum. The final spectra are zoomed in the velocity 
range to a little bit wider than CO 1-0 line width from literature, and a zero level baseline 
is subtracted to get the `local' baseline level. We smoothed the spectra 
to  resolution from $\sim$20 to 50 km s$^{-1}$, depending on the 
noise levels, and  convert from an antenna temperature, T$_A$, to the
 main-beam temperature, T$_{mb}$, by	 scaling the main-beam efficiency, 
 $\eta_{mb}$,  using $T_{mb} = {T_A}^\star/{\eta _{mb}}$.  Then  the CS 5-4 
line flux in each galaxy was  integrated  with the velocity range  of CO line from literature, and listed as the second column in Table 2, including the integrated flux and error bar.

Single component gaussian profile  is also used to fit  each of the  spectra
with solid detections, to derive the area, system velocity, and 
line width of these lines.  The fitted line flux of each galaxy was listed in the third column, while the fourth  column was the line width. 

   The upper limits of the CS measurements correspond to sources without solid  
 detections,  where  3 $\sigma$   are used 
in the following analysis. The upper limits were calculated through the relation
\begin{equation}\label{eq-upper}
T_a \Delta \rm{v} \;\; (\rm{K \; km\,s^{-1} })\;\; \frac{3 T_a^{rms} 
\Delta \rm{v}}{\sqrt{\Delta \rm{v}/\Delta \rm{v}_{res}}}  
\end{equation}
where $\Delta \rm{v}$ is the FWZI (Full Width Zero Intensity) line width of galaxies;
$\Delta \rm{v}_{res}$ is the velocity resolution (in our case $\approx 
20 - 50$ km\,s$^{-1}$  after smoothing) and 
$T_a^{rms}$ is from the baseline fitting.

\subsection{CS line luminosity and the infrared luminosity correction}

Most of the dense gas emission in  galaxies with distance larger than that of NGC 2276 would be expected to concentrate 
within the SMT beam ($\sim$ 33$\arcsec$ for CS 5-4).  We would like to consider that we have picked up  most of the CS 5-4 emission and will compare with the total infrared luminosity of entire galaxy in those sources.    On the other hand, infrared luminosity in  6 sources from NGC 6946 to NGC 2276 at the beginning of  Table 1 have been corrected  from the whole galaxy to the region within SMT beam. We used MIPS 24$\mu$m image of those galaxies to obtain the ratio of such region to the whole galaxy and then derived infrared luminosity with CS 5-4 beam by scaling  the  total infrared luminosity listed in Table 1 from \cite{2003AJ....126.1607S} with 24$\mu$m flux ratio. 

We computed the CS line luminosity using 

\begin{equation}
 L'_{CS} = \int T_b(CS) dV d\Omega_s d^2_A \approx \pi/(4 ln 2) \theta_{mb}^2 I_{CS} d_L^2 (1+z)^{-3}
\end{equation},
where $I_{CS}$ is defined as:

\begin{equation}
I_{CS} = \int T_{mb}(CS) d V = \int T_b(CS) dV d\Omega_s/[(1+z)\Omega_{s\star b}]
\end{equation}, where    $L'_{CS}$ is in the unit of $ K ~km~ s^{-1} pc^2$.   These 2 formulae were the formulae (1) and (2)  in \cite{2004ApJS..152...63G}, where we  used CS 5-4 instead of HCN 1-0.  
All of the measured line fluxes along with their errors are tabulated in Table 2.
The errors include thermal and calibration errors,  and also errors on the
assumed aperture efficiencies. 
Each of these is assumed to be the uncorrelated and thus is added in the
quadrature. The noise level was calculated for each line   using $RMS_{line} =RMS_{channel}\times\sqrt{N_{channel}\times FWZI}$,
assuming all channels have the same response and are without any intereference between
each other.

\section{Results and discussion}

We detected  CS 5-4 with velocity integrated flux higher  than 3$\sigma$ level in 13 galaxies. We used the CO line width to obtain  velocity integrated flux,  except for NGC 6240  in which we used CS 7-6 line range from    \cite{2009ApJ...692.1432G}. Since CO line width  is wider than that of dense gas tracers such as CS 5-4, we may over-estimate the noise level  and all  the  13 sources with velocity integrated flux higher  than 3$\sigma$ should be solid detections. 
 We show  spectra of 2 detected galaxies  in the top (Arp 220) and bottom (NGC 1068) of  Figure 1.  CS 2-1, 3-2, 5-4 and 7-6 lines in Arp 220  have been observed by  \cite{2009ApJ...692.1432G}  and showed different line shapes for different transitions.  The velocity range and total intensity of our CS 5-4  spectrum    agree well  with the result from \cite{2009ApJ...692.1432G}, but with a  little bit difference in the line shape. Our CS 5-4 spectrum in NGC 6240  shows a similar velocity range to that of  CS 7-6  observed by \cite{2009ApJ...692.1432G}, which is about 200-300 km/s  blue shifted than CO lines. 
Even though the integrated flux of  IRAS 23365+3604   is less than 3$\sigma$,   the peak intensity at the velocity resolution of 23 km/s is  more than   5mK, which is about 4$\sigma$ as shown at the top of Fig 2.  We consider that CS 5-4 emission is detected in IRAS 23365+3604  and the flux derived from  Gaussian fitting was used.

To avoid the system uncertainty between different telescopes, we have not included the CS 5-4 data detected with other telescopes for our study. The fit results for    14 sources with CS 5-4 detected sources is: $Log(L_{IR})(L_{\odot})=0.938(\pm0.1)\times Log(L_{CS})(K~km~s^{-1}\rm{pc}^{-1})+4.156(\pm0.834)$, with   correlation coefficient of   0.939. 
 NGC 2146 and NGC 1068 have the largest deviations from the fit as shown  in Figure 3.  NGC 1068 is known to harbor active galactic nucleus with non-negligible  contribution to the total 24 $\mu$m emission. So the   infrared luminosity within the CS 5-4 beam    using the aperture correction of the  24$\mu$m MIPS image  could be over-estimated, and thus NGC 1068 is likely closer to the best fit. Our   CS 5-4 spectrum has similar line shape and velocity range to those  obtained with IRAM 30m \citep{2009ApJ...694..610M}, but with  total flux  about twice  that from IRAM 30M.  Since the beam size of IRAM 30m, which is  about 10$''$ for observing CS 5-4,   is much smaller than the CS emission region, which is about 15$''$ from  high resolution observation of  CS 2-1 with IRAM PdBI \citep{1997Ap&SS.248...59T}. So our observation with HHSMT can pick up more extended emission than that with IRAM 30m.  On the other hand,  as a nearby LIRG with only a possible low luminosity AGN \citep{2005PASJ...57..135I, 2009A&A...506..689I}, the infrared luminosity of  NGC 2146 within CS 5-4 HHSMT beam should not be over estimated as that for NGC 1068.  The large deviation should not be from  the offset problem of  beam switching mode, because the 2$'$ offset used in the observation is much far from the disk region with molecular gas based on CO mapping observation  \citep{2009A&A...506..689I}.      Because of limited velocity coverage and  broad line of CS 5-4    (see Figure 2) with the full line width of about 700 km/s, which is constant with  the result of HCN 1-0  \citep{2004ApJS..152...63G}, the over subtracted  baseline of CS 5-4  may cause under estimation of the line flux, which can cause the large deviation.  Since it is hard to do re-calibration of CS 5-4 and/or infrared emission in NGC 2146 and NGC 1068, we will leave these two sources in this sample as solid detection and do fitting with original CS and infrared luminosity.  We also tried to do the same fitting but without   NGC 2146 and NGC 1068, which gave the result:  $Log(L_{IR})(L_{\odot})=1.037(\pm0.06)\times Log(L_{CS})(K~km~s^{-1}\rm{pc}^{-1})+3.29(\pm0.50)$, with   correlation coefficient of   0.981.

The relationship between CS 5-4 and infrared luminosities presented
in this work is consistent with close associations between star
formation and dense gas which is also suggested by previous studies
of correlations between SFR and HCN 1-0 \citep{1992ApJ...387L..55S, 2004ApJ...606..271G}.   With  energy from upper level to the grand level  $E_{up}\sim 35K$, which is much higher than the temperature ($ 10-20K$) of quiescent  molecular cloud without star formation, and  a little bit higher effective critical density than that of HCN 3-2 which is about 10 times  that for HCN 1-0 \citep{1999ARA&A..37..311E}, CS 5-4 is  a good  tracer of active star forming  gas. Dense gas mass in galaxies estimated from individual types of  dense gas tracers, such as HCN 1-0 and  CS 5-4, may have systematic errors due to different physical conditions (temperature, density, etc.)  excitation mechanisms (infrared pumping instead of collisional
excitation) in different galaxies. .   The contribution of nuclear activity to the total infrared luminosity can cause the over estimation of star formation rate, which can also introduce   scatter to the correlation.  Even though the line ratios of different dense gas  tracers have been found to be variable in different galaxies  \citep{2002A&A...381..783A, 2009ApJ...692.1432G},  the relation between  dense gas mass, traced by both HCN 1-0 \citep{2004ApJ...606..271G} and CS 5-4 in this work, and the star formation rate traced by infrared luminosity has an unity  slope.    
So we suggest that dense gas, which is directly linked to star formation,   is very important  for understanding star formation in galaxies, and the relation between star formation and dense gas should be similar with the observations of different dense gas tracers,  otherwise  that dense gas tracer should  be affected by other properties, such as shock or infrared pumping.
Recent  theoretical  work  \citep{2009ApJ...699..850K}   showed that both the total gas and the molecular gas star formation law can not even  be fitted by a single  power law.  Theoretical predictions  of dense gas star formation law  and    dense gas fraction in  different type of galaxies  are  needed  for comparing with  observational results.

 \section{Summary and prospects}

With the observations of CS 5-4 using the HHSMT 10m telescope toward 24 local infrared bright galaxies including 20 LIRGs/ULIRGs and 4 nearby normal galaxies, we detected CS 5-4 emission in 12 LIRGs/ULIRGs and 2  nearby normal galaxies. As the first extragalactic  CS 5-4 survey with a sample larger than 10 galaxies, we  found a correlation between CS 5-4 luminosity and infrared luminosity with an  almost unity slope. 
  
Since   only 14 galaxies have CS 5-4 detection in our sample, it is necessary to enlarge the sample  for CS 5-4 observation.  The existing (sub-)millimeter telescopes, such as HHSMT, CSO, JCMT, IRAM 30m, APEX, and also the LMT  can be used for such observations.   The nearby star forming galaxies can be observed with the smaller telescopes (HHSMT,  CSO, JCMT, and APEX), while the ULIRGs/LIRGs should be observed with IRAM 30m and the LMT which have better sensitivity.  We have started a project to detect CS 5-4 in  more ULIRGs/LIRGs with IRAM 30m. Other CS transitions, such as CS 1-0, 2-1, 3-2, and also 7-6 can also be considered to determine dense gas properties (volume density, temperature, etc.) for galaxies with strong CS emission based on one transition observation.  CS lines in galaxies should also be a good  choice to study star formation in galaxies for  ALMA early science.

\section*{Acknowledgements}

The authors thank the HHSMT  staff   for   the help of remote observation and thank the anonymous referee for helpful comments.  This work was   supported   by  the Natural Science Foundation of China under  grants of   10803002  and 10833006.

 \bibliographystyle{mn2e}
 \bibliography{CS54_MN}

 \clearpage{}





\begin{table} 
\begin{center}
\caption{CS 5-4 survey source list and galaxy properties\label{table-3}}
\tiny
\begin{tabular}{llllllllllll}
\hline
Source name      & R.A.        &  DEC       &D   &Redshift        &Log($L_{IR}$)      \\
                 & (J2000)     &(J2000)       &Mpc     &          &                             $L_{\odot}$      \\
\hline
NGC~6946         & 20:34:52.3  &+60:09:14     &3.65    &0.000133        & 10.16    \\
NGC~891          & 02:22:33.4  &+42:20:57     &4.43    &0.001761        & 10.27      \\
NGC~2146         & 06:18:37.6  &+78:21:19     &12.6    &0.002979       & 11.07    \\
NGC~7331         & 22:37:04.1  &+34:24:56     &14.5    &0.002722       &   10.58            \\
NGC~1068         & 02:42:40.7  &-00:00:48     &16.1    &0.003793      & 11.27          \\
NGC~2276         & 07:27:14.3  &+85:45:16     &33.8    &0.008059      &  10.81           \\ 
NGC~4194         & 12:14:09.5  &+54:31:37     &35.5    &0.008342        & 11.06      \\
CGCG~049-057     & 15:13:13.1  &+07:13:32     &55.5    &0.012999        & 11.27    \\
NGC~992          & 02:37:25.5  &+21:06:03     &55.7    &0.013813  &           11.02      \\
IRAS~17578-0400  & 18:00:31.9  &-04:00:53     &60.0    &0.014043  &   11.35        \\
NGC~7771         & 23:51:24.9  &+20:06:43     &60.9    &0.014267  & 11.34           \\
MCG+12-02-001    & 00:54:03.6  &+73:05:12     &67.1    &0.015698  &  11.44  \\
NGC~1614         & 04:33:59.9  &-08:34:44     &68.1    &0.015938  & 11.6        \\
NGC~7469         & 23:03:15.6  &+08:52:26     &69.8    &0.016317  &  11.59  \\
NGC~828          & 02:10:09.6  &+39:11:25     &76.8    &0.017926  &  11.31  \\
Arp~220          & 15:34:57.1  &+23:30:11     &77.6    &0.018126  &  12.21   \\
NGC~2623         & 08:38:24.1  &+25:45:17     &79.3    &0.018509  &  11.54   \\
VV~114           & 01:07:47.2  &-17:30:25     &86.1    &0.020067  &  11.65   \\
NGC~6240         & 16:52:58.9  &+02:24:03     &106     &0.024480  &  11.85       \\
NGC~6090         & 16:11:40.7  &+52:27:24     &126.6   &0.029304  &  11.51  \\
Mrk~231          & 12:56:14.2  &+56:52:26     &184.0   &0.042170  &  12.51  \\
IRAS~17208-0014  & 17:23:21.9  &-00:17:00     &187.0   &0.042810  & 12.39 \\
VII~zw31         & 05:08:15.3  &+79:36:46     &236.1   &0.053670  &  11.94   \\
IRAS~23365+3604  & 23:39:01.3  &+36:21:09     &286.0   &0.064480  &  12.13  \\    
\hline
\end{tabular}
\end{center}
 $a.$ Total Infrared Luminosity is from RBGs (Sanders 2003), and corrected with the distances.\\
 $b.$  The sources from NGC 6946 to NGC 2276 have been corrected from total infrared luminosity from this table to the emission within the  region of CS 5-4 beam, using MIPS 24 $\mu$m image, to compare with CS 5-4 luminosity. 

\end{table}

\begin{table} 
\begin{center}
\caption{Results of CS 5-4 line observation with HHSMT}
\tiny
\begin{tabular}{llllllllllll}
\hline \hline 
Source name       &    I$\rm_{CS 5-4}$                     &Area(CS5-4)     &$\Delta$V$\rm_{CS 5-4}$            &On source time\\   
                  &    K km/s                     &   K km/s          &    km/s                  &Minutes   \\    
\hline
NGC~6946          &     1.299 $\pm$ 0.097         & 1.40$\pm$0.19  &480 $\pm$70               &45    \\ 
NGC~891           &     3.692 $\pm$ 0.327         & 3.64$\pm$0.50  &377 $\pm$54               &20    \\ 
NGC~2146          &     0.611 $\pm$ 0.060        & 0.60$\pm$0.07  &315 $\pm$54               &110    \\ 
NGC~7331          &$<$  0.626 $\pm$ 0.266        & ----           &  ----                    &15    \\ 
NGC~1068          &     0.883 $\pm$ 0.180         & 0.67$\pm$0.14  &130 $\pm$29               &60    \\ 
NGC~2276          & $<$ 0.72  $\pm$ 0.361      & ----           &  ----                    &15    \\ 
NGC~4194          &     0.976 $\pm$ 0.298        & ----           &  ----                    &45    \\ 
CGCG~049-057      &     1.100 $\pm$ 0.304         & 1.11$\pm$0.43  &304$\pm$105               &45    \\ 
NGC~992           &$<$  0.138 $\pm$ 0.159       & ----           &  ----                    &30    \\ 
IRAS~17578-0400   &     0.908 $\pm$ 0.133         & 0.90$\pm$0.16  &283$\pm$50                &45    \\ 
NGC~7771          &     0.419 $\pm$ 0.124      & 0.46$\pm$0.16  &296$\pm$88                &60    \\ 
MCG+12-02-001     &     0.642 $\pm$ 0.060       & 0.69$\pm$0.08  &368$\pm$41                &70    \\ 
NGC~1614          &$<$  0.088 $\pm$ 0.044     & ----           &  ----                    &140    \\ 
NGC~7469          & $<$ 0.244 $\pm$ 0.167        & ----           &  ----                    &40    \\ 
NGC~828           &$<$  0.202 $\pm$ 0.124        & ----           &  ----                    &20    \\ 
Arp~220           &     1.631 $\pm$ 0.128        & 1.70$\pm$0.22  &467$\pm$68                &60    \\ 
NGC~2623          &$<$  0.147 $\pm$ 0.083        & ----           &  ----                    &100    \\ 
VV~114            &$<$ -0.441 $\pm$ 0.249         & ----           &  ----                    &15    \\ 
NGC~6240          &     0.993 $\pm$ 0.118       & 1.06$\pm$0.18  &289$\pm$58                &80    \\ 
NGC~6090          &$<$  0.110 $\pm$ 0.160        & ----           &  ----                    &100    \\ 
Mrk~231           &$<$ -0.390 $\pm$ 0.117         & ----           &  ----                    &45    \\ 
IRAS~17208-0014   &     0.981 $\pm$ 0.330        & 0.86$\pm$0.32  &141$\pm$65                &20    \\ 
VII~zw31          &     0.393 $\pm$ 0.109       & 0.46$\pm$0.12  &341$\pm$81                &140    \\ 
IRAS~23365+3604   &$<$  0.309 $\pm$ 0.177        & 0.32$\pm$0.09  &64$\pm$25                 &60    \\ 
\hline
\end{tabular}
\end{center}
 $a.$ I$\rm_{CS 5-4}$   is the flux  integrated from the spectrum with the line width of CO 1-0 from literature.\\
 $b.$ Area(CS5-4) is the flux using  gaussian fitting for sources with solid detection except for NGC 4194, which has the line shape more complicated than one gaussian component. The derived line width was listed as 
$\Delta$V$\rm_{CS 5-4}$.

\end{table}

 \clearpage

\begin{figure}
\includegraphics[angle=-90,scale=.30]{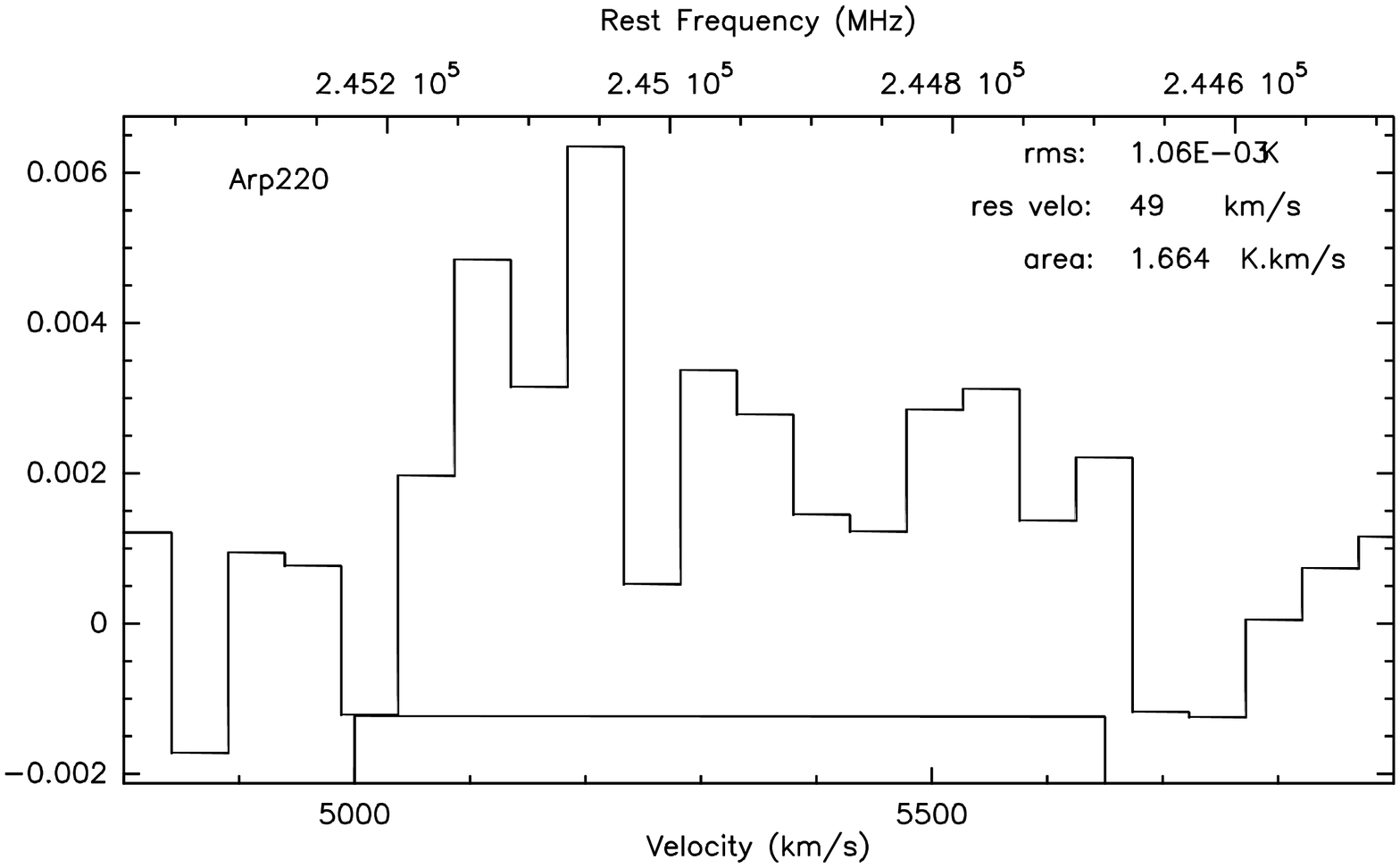}

\includegraphics[angle=-90,scale=.30]{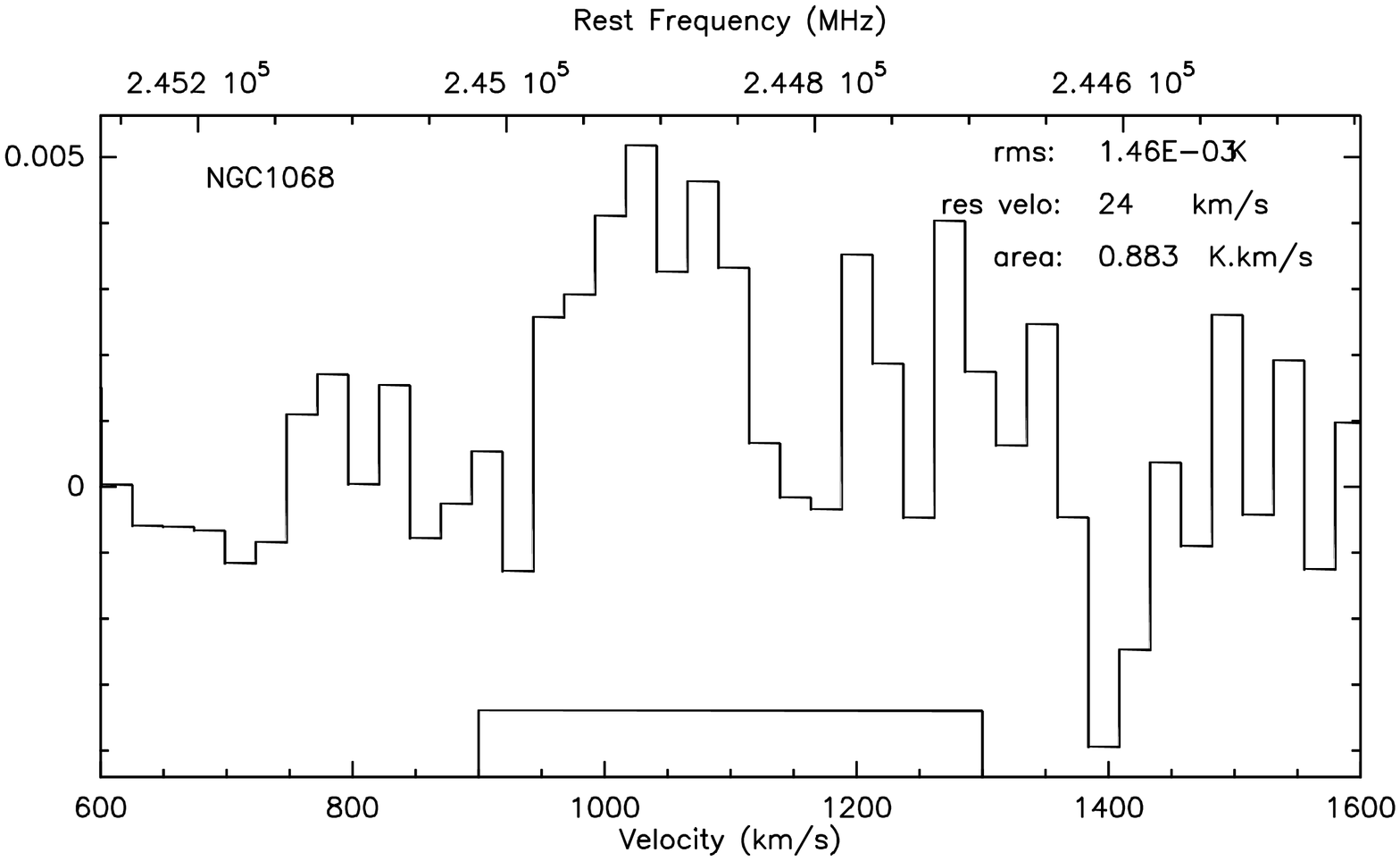}

\caption{CS 5-4 spectra of Arp 220 (top) and NGC 1068 (bottom)  observed with HHSMT.}
\end{figure}

\begin{figure}

\includegraphics[angle=-90,scale=.30]{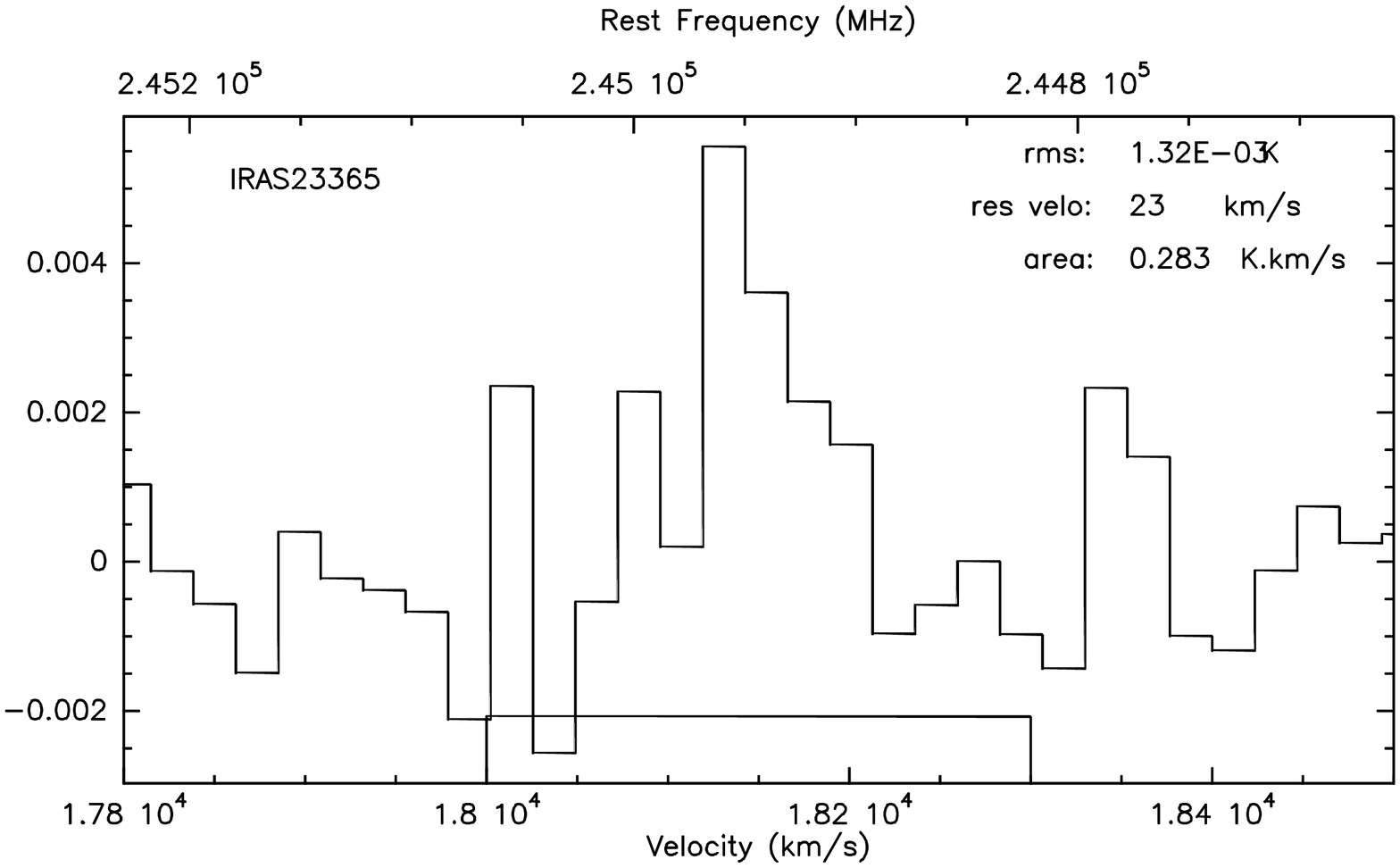}
\includegraphics[angle=-90,scale=.30]{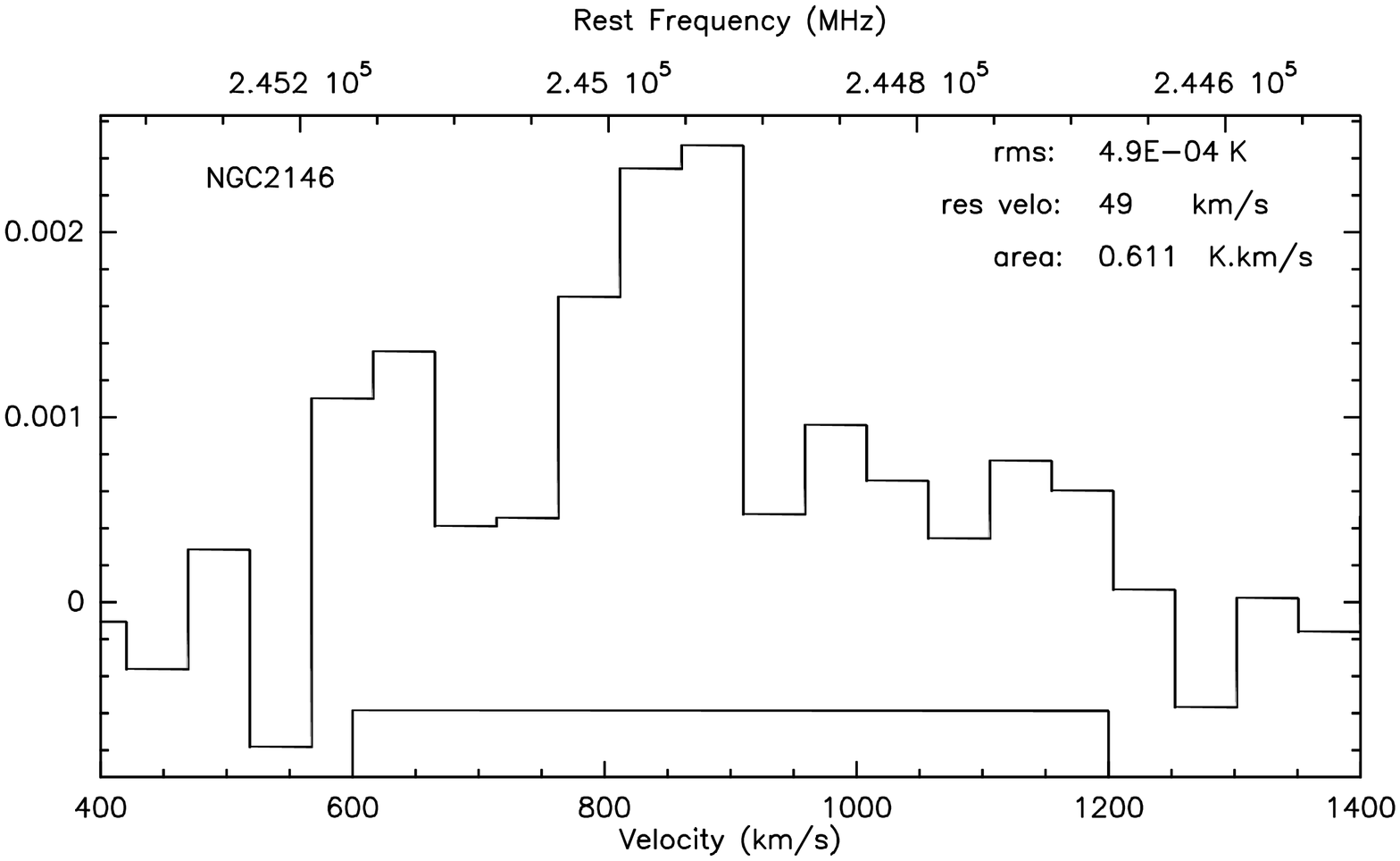}
\caption{CS 5-4 spectra of IRAS23365+3604 (top) and NGC 2146 (bottom)  observed with HHSMT. }
\end{figure}

\begin{figure}
\includegraphics[angle=90,scale=.30]{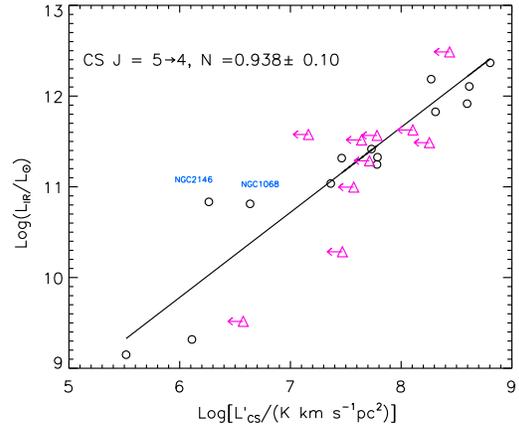}

\caption{The 14 galaxies with CS 5-4 detection are plotted  as `$\circ$', while the other 10 sources are plotted with left arrow and `$\diamond$' using the 3$\sigma$ of the integrated flux to estimate the upper limit of  CS 5-4 luminosity.  
 }
\end{figure}

 

  \label{lastpage}

 \clearpage
\end{document}